\documentclass[manuscript,screen]{acmart}

\usepackage{cleveref}
\usepackage[nolist]{acronym}
\usepackage{paralist}
\usepackage{hyperref}

\usepackage{multirow}
\usepackage{pbox}
\usepackage{colortbl}
\usepackage{hhline}

\usepackage[hyphenbreaks]{breakurl}
\usepackage{url}

\usepackage{color}

\newcommand{\quotes}[1]{`#1'}

\acmBooktitle{Proceedings}

\setcopyright{none}
\renewcommand\footnotetextcopyrightpermission[1]{} 
\begin{document}

\title{Foreseeing the Impact of the Proposed AI Act on the Sustainability and Safety of Critical Infrastructures}
\renewcommand{\shorttitle}{Foreseeing the Impact of the Proposed AI Act}

\author{Francesco Sovrano}
\authornote{Both authors contributed equally to this research.}
\affiliation{%
  \institution{Department of Computer Science and Engineering, University of Bologna}
  \streetaddress{Via Zamboni 33}
  \city{Bologna}
  \country{Italy}
}
\email{francesco.sovrano2@unibo.it}

\author{Giulio Masetti}
\authornotemark[1]
\affiliation{%
  \institution{Istituto di Scienza e Tecnologia dell'Informazione - Consiglio Nazionale delle Ricerche}
  \streetaddress{Via Moruzzi 1}
  \city{Pisa}
  \country{Italy}}
\email{giulio.masetti@isti.cnr.it}


\begin{abstract}
    The AI Act has been recently proposed by the European Commission to regulate the use of AI in the EU, especially  on high-risk applications, i.e. systems intended to be used as safety components in the management and operation of road traffic and the supply of water, gas, heating and electricity.
    On the other hand, IEC 61508, one of the most adopted international standards for safety-critical electronic components, seem to mostly forbid the use of AI in such systems.
    Given this conflict between IEC 61508 and the proposed AI Act, also stressed by the fact that IEC 61508 is not an harmonised European standard, with the present paper we study and analyse what is going to happen to industry after the entry into force of the AI Act.
    In particular, we focus on how the proposed AI Act might positively impact on the sustainability of critical infrastructures by allowing the use of AI on an industry where it was previously forbidden.
    To do so, we provide several examples of AI-based solutions falling under the umbrella of IEC 61508 that might have a positive impact on sustainability in alignment with the current long-term goals of the EU and the Sustainable Development Goals of the United Nations, i.e., affordable and clean energy, sustainable cities and communities.
\end{abstract}



\keywords{AI Act, IEC 61508, safety standards, sustainability}

\maketitle

\begin{acronym}	
	\acro{SIL}{Safety Integrity Level}
	\acro{AI}{Artificial Inteligence}
	\acro{ML}{Machine Learning}
	\acro{AIA}{\acs{AI} Act}
	\acro{SC}{\emph{full compliance with international standards}}
	\acro{SA}{\emph{state of art}}
	\acro{SE}{\emph{safety which consumers may reasonably expect}}
	\acro{EU}{European Union}
\end{acronym}

\section{Introduction} \label{sec:intro}

Harnessing the full potential of AI could lead our society to more efficient and thus sustainable energy production, storage and transportation, in accordance with many of the Sustainable Development Goals of the United Nations \cite{nations2016sustainable}, including: 
\begin{inparadesc}
    \item affordable and clean energy (goal 7),
    \item industry, innovation and infrastructure (goal 9),
    \item sustainable cities and communities (goal 11),
    \item responsible consumption and production (goal 12),
    \item climate action (goal 13).
\end{inparadesc}
Indeed, AI can be used to optimally recognise, predict, detect, identify, determine, control, generate, and classify \cite{munro2019jobs} in a wide range of tasks, sometimes also achieving or exceeding human performance in problems such as strategy games \cite{gibney2016google,schrittwieser2020mastering}, image and object recognition \cite{he2015delving,rajpurkar2017chexnet}, etc.
Nonetheless, the adoption of AI-based technological solutions for more sustainable energy (e.g., for decreasing the carbon emissions of coal-fired thermal power plants \cite{sivageerthi2022modelling}) has been held back in the last decades by conservative international standards (i.e., IEC 61508 \cite{smith2020safety}: a standard that regulates safety-critical electronic components and that practically forbid AI in many critical infrastructures).

Despite this, in 2021, the European Commission published a proposal of AI Act\footnote{EUR-Lex - 52021PC0206 - EN - EUR-Lex} \cite{AIA} that is expected to become a legally binding regulation to all the Member States of the EU by 2024.
Importantly, the objective of the AI Act is to set a common regulatory and legal framework for AI that applies to all sectors (except for military), and to all types of artificial intelligence, including (high-risk) AI for the \textit{management and operation of critical infrastructure}.
Considering that one of the goals of the proposed AI Act is to regulate the use of AI also on those systems covered by IEC 61508, i.e. \quotes{systems intended to be used as safety components in the management and operation of road traffic and the supply of water, gas, heating and electricity} (see Annex III, point 2.a), the research questions we are trying to answer with the present paper are the following:
\begin{itemize}
    \item Will the AI Act be disruptive with respect to IEC 61508? 
    \item What will happen to those industries currently regulated by IEC 61508?
\end{itemize}
In fact, we believe that answering these has the potential to help both industry and academia to quickly seize the opportunities offered by the new European policies enshrined in the AI Act.


In order to answer these questions, we analyse the main differences between IEC 61508 and the proposed AI Act.
Then, we identify significant and concrete examples of technical solutions that could increase sustainability and energy efficiency but which, at the moment, are not feasible due to IEC 61508.
Furthermore, we also study how such new technological solutions might impact industry, trying to understand how disruptive the AI Act could be by loosening the tight laces of IEC 61508 on AI.
Hence, we try to align our findings to the medium- and long-term objectives of the EU on sustainability and support for innovation.

This paper is structured as follows.
Section \ref{sec:met} discusses the adopted methodology.
In Section \ref{sec:background} we give enough background to understand IEC 61508 and its implications for industry.
While, in Section \ref{sec:analysis} we analyse the position of the AI Act on IEC 61508 and other standards, providing in Section \ref{sec:classification} our understanding of how AI could improve sustainability and energy efficiency whilst maintaining safety.
Finally, in Section \ref{sec:conclusion} we try to give a conclusive answer to each research question, discussing the consequences of our findings as well as some possible issues.

\section{Methodology} \label{sec:met}

The adopted methodology employed for analysing the proposed AI Act and answering the aforementioned research questions is as follows.
First of all, we start from the identification of climate-neutrality (by 2050) as one of the main and most actual objectives of the EU.
In particular, we refer to the European Green Deal\footnote{\url{https://eur-lex.europa.eu/legal-content/EN/TXT/?uri=COM:2019:640:FIN}} and the EU’s commitment to global climate action under the Paris Agreement, considering them as interpretative keys for the proposed AI Act.
Hence, we study how articles such as article 54.1.a, interpreted under the lenses of the European Green Deal, may impact on the adoption of new AI-based technologies in industrial contexts currently regulated by non-harmonised technical standard (i.e., IEC 61508) that practically forbid the use of AI.
We do it by looking at how the AI Act may help to improve the sustainability of our society thanks to state-of-the-art advancements in AI that cannot currenlty be deployed.

\section{Background} \label{sec:background}
With this section we provide a minimal amount of information about safety, AI, the IEC 61508 standard and the proposed AI Act.

\subsection{The safety standard IEC 61508}
\label{whatIsIEC61508}
IEC 61508 \cite{IEC61508} is an international standard describing how to design, deploy 
and maintain an Electrical or (Programmable) Electronic safety-related system.
Examples of safety-related systems to which IEC 61508 can be applied are:
\begin{inparadesc}
    \item emergency shut-down systems,
    \item remote monitoring, 
    \item operation or programming of a network-enabled process plant,
    \item information-based decision support tool where erroneous results may affect safety. 
\end{inparadesc}
In particular, programmable electronic safety-related systems typically incorporate programmable controllers, programmable logic controllers, microprocessors, application specific integrated circuits, or other programmable devices (e.g., \quotes{smart} devices such as sensors/transmitters/actuators).
The focus is in particular on safety functions and on 
the relative level of risk reduction that they provide.
Those levels are grouped in four \acfp{SIL}, the higher the \acl{SIL}
the greater the risk of failure.


Notice that it is expected\footnote{\url{https://www.iec.ch/functional-safety/faq}} 
that IEC 61508 can be published as EN 61508, an European standard, but it does not have the status of a \textit{harmonized} European standard in relation to any EC product directive and it is not therefore listed in the EC Official Journal. 
However, this does not prevent compliance with relevant parts of EN 61508 being used to support a declaration of conformity with an EC product directive, if that is appropriate. 
In any cases, IEC 61508 is followed
worldwide\footnote{\url{https://www.iec.ch/national-committees}}.


\subsection{Safety vs AI}
\label{safety_vs_AI}

Quoting \cite{SafetyHandbook}: \quotes{
    There is no such thing as zero risk. 
    This is because no physical item has zero failure rate, no human being makes zero errors, and no piece of software design can foresee every operational possibility
}.
Thus, perfect \emph{safety}, i.e., the absence of catastrophic consequences on the user(s) and the environment \cite{AvizienisTaxonomy}, is out of reach.
During the last decades, several standards on how to develop hardware and software artifacts in safety critical contexts have been defined.
These standards crystallize lessons learned, common practices and scientific research into concrete guidelines. 
Each industry sector has its own standard, but the idea behind all of them is the same: a risk-based approach that characterize the entire product life-cycle.

With respect to \ac{AI}, at row 5 in Tables A.2 and C.2, Part 3, of IEC 61508 \cite{IEC61508}, it is clearly stated that \ac{AI} is not recommended for \acl{SIL} 2 or above because it may complicate the achievement of one or more of the following properties: 
\begin{inparadesc}
    \item correctness with respect to software safety requirements specification,
    \item freedom from intrinsic design faults,
    \item simplicity and understandability, 
    related with the observability-in-depth principle,
    aimed at avoiding as much as possible a false sense of safety due to lack of information,
    \item predictability of behaviour,
    \item verifiable and testable design.
\end{inparadesc}

IEC 61508 has influenced other standards \cite{SafetyHandbook},
here called `second tier standards', 
that are as rigid as IEC 61508 Part 3 with respect to \ac{AI}.
Among those, examples are software for train EN 50128 \cite{EN50128},
process industry \cite{IEC61511} 
and machinery IEC 62061 \cite{IEC62061}.
Parallel to the family of standards originated from IEC 61508,
other really important examples where \ac{AI} is banned, for high \acl{SIL}, from 
computer-based systems employed in nuclear power plants, IEC 60880 \cite{IEC60880}, 
and avionic, DO-178 C \cite{DO-178C}.

To the best of the authors' knowledge, the only safety standard
that allows the employment of \ac{AI} (because it does not mention it, and then
it is not `not recommended') is ISO 26262 for the automotive industry sector
\cite{GapISO26262andML,VetVforMLinISO26262,salay2017MLinISO26262}.

\subsection{The Proposed AI Act} \label{sec:proposed_ai_act}

The AI Act \cite{AIA} is a proposed European law on AI.
Differently from other domains, this act is specific to AI systems and requires an \textit{ad hoc} discussion rather than the framing of these systems in the discussion of other legal domains. 
This is because AI technologies are not placed within an existing legal framework (e.g., banking), but the whole legal framework (i.e., the proposed AI Act) is built around AI technologies. 

The proposed AI Act assigns applications of AI to three risk categories. 
First, applications and systems that create an unacceptable risk, such as government-run social scoring of the type used in China, are banned. Second, high-risk applications, such as a CV-scanning tool that ranks job applicants, are subject to specific legal requirements. Lastly, applications not explicitly banned or listed as high-risk are largely left unregulated.

Examples of high-risk AI are given by the proposed AI Act in Annex III, as they broadly include applications for:
\begin{inparadesc}
    \item biometric identification and categorisation of natural persons,
    \item management and operation of critical infrastructures,
    \item etc.
\end{inparadesc}
In particular, for all those applications defined as \quotes{high-risk}, the AI Act provides several limitations and safety assurance procedures including:
\begin{inparadesc}
    \item a risk management system (art. 9),
    \item appropriate data governance and management practices (art. 10),
    \item detailed technical documentation (art. 11).
\end{inparadesc}


Finally, the AI Act defines in Annex I what are the AI techniques and approaches referred by the proposal.
Among them we have: 
\begin{inparadesc}
    \item machine learning approaches (e.g., neural networks),
    \item logic- and knowledge-based approaches (e.g., inductive logic programming),
    \item and statistical approaches (e.g., Bayesian estimation).
\end{inparadesc}

\section{Analysis of the Position of the AI Act on IEC 61508 and Other Standards} \label{sec:analysis}

It is crystal clear from the European Green Deal and the EU’s commitment to global climate action under the Paris Agreement that one of the big goals of the EU is to be climate-neutral by 2050.
Citing the words of the European Commission: \quotes{The EU can lead the way [to climate-neutrality] by investing into realistic technological solutions, empowering citizens and aligning action in key areas such as industrial policy, finance and research, while ensuring social fairness for a just transition.}
The reason why we are citing these statements is that we are going to use them as interpretative key for the proposed AI Act, especially with respect to the importance of article 54.1.a, stating that \quotes{innovative AI systems shall be developed for safeguarding substantial public interest in [...] a high level of protection and improvement of the quality of the environment}.

In fact, the AI Act is (as mentioned in Section \ref{sec:background}) regulating a vast range of AI applications, with due focus on those listed as high-risk in Annex III.
In particular, it covers, among others, the AI applications for the \quotes{management and operation of critical infrastructure}, i.e. the \quotes{AI systems intended to be used as safety components in the management and operation of road traffic and the supply of water, gas, heating and electricity.}
But, considering that many critical infrastructures are currently following a non-harmonized IEC 61508 standard that is de facto excluding the involvement of any AI, we can see a non-alignment of it to the proposed AI Act.

Indeed, according to article 40 only the \quotes{harmonised standards or parts thereof the references of which have been published in the Official Journal of the European Union} are considered to be in conformity with the requirements set out in the proposed AI Act for high-risk AI systems (see Chapter 2 of Title III).
In other words, article 40, together with the \textit{Explanatory Memorandum}, article 54.1.a and the fact that IEC 61508 is a non-harmonised standard, make us understand that the intent of the proposed AI Act is to promote innovative AI systems also for the \quotes{management and operation of road traffic and the supply of water, gas, heating and electricity}.
As consequence, we envisage that the proposed AI Act, without further modifications, can have a disruptive effect in the industry of critical infrastructures.
This effect can be disruptive in a positive way, by opening to new technological solutions that have the potential to improve even further our quality of life, reducing costs and increasing efficiency.
Nonetheless, it can be disruptive also in a negative way, by ceding the control of critical infrastructures to automatic decision makers that are possibly opaque, greedy, unfair and non-transparent in a way that would not allow to understand where the responsibility lies.

Although, despite the fact that IEC 61508 is a non-harmonized standard, thus not covered by article 40, we can see that the proposed AI Act shares several and important similarities with it, suggesting that it is not the intent of the EU Commission to fully upset existing standards.

\begin{table}[!tb]
\centering
\caption{\textbf{AI Act vs IEC 61508}: AI Act is centred on transparency while IEC 61508 on safety. This table shows how the proposed AI Act and IEC 61508 address the same process for risk-assessment, analysis, development and document production in different ways.}
\label{IEC61508andAIAsimilarities}
\makebox[\linewidth]{
\begin{tabular}{|p{0.3\linewidth}|p{0.3\linewidth}|p{0.3\linewidth}|}
\hhline{~--|}
\multicolumn{1}{c|}{\multirow{2}{*}{}}                                                                                      & \multicolumn{2}{c|}{{\cellcolor[rgb]{0.918,0.918,0.918}}\textbf{Differences}}                                                                                                                                                                             \\ 
\hhline{~--|}
\multicolumn{1}{c|}{}                                                                                                       & {\cellcolor[rgb]{0.918,0.918,0.918}}\textbf{IEC 61508}                                                                   & {\cellcolor[rgb]{0.918,0.918,0.918}}\textbf{Proposed AI Act}                                                          \\ 
\hline
{\cellcolor[rgb]{0.918,0.918,0.918}}Risk-based approach, in particular to establish the belongings to a predefined category & Quantitative (hazard analysis, risk assessment and identify the \acl{SIL})                                  & Qualitative (one of the alternatives: no risk, application listed in Annex III, AI not applicable)                    \\ 
\hline
{\cellcolor[rgb]{0.918,0.918,0.918}}Normalised life cycle, with focus on accountability                                   & V-shape development (focus on modularity and decomposability)                                                            & Clear definition of datasets (focus on data management, in particular for training the AI and how to use the product  \\ 
\hline
{\cellcolor[rgb]{0.918,0.918,0.918}}Ex-ante and ex-post analysis                                                            & Statistical methods (hardware), study of qualitative techniques (hardware and software) and structured testing campaigns & Declarative (identify high level characteristics, provide general description of components behaviour)                \\ 
\hline
{\cellcolor[rgb]{0.918,0.918,0.918}}Document production                                                                     & Assessment performed by an external institution                                                                          & Fill a form in EU database, part of the information is of public domain (focus on transparency)                       \\
\hline
\end{tabular}
}
\end{table}

Overall, we see that the intent of the proposed AI Act is to modernize existing critical infrastructures, to make them more sustainable.
To do so, the AI Act does not ignore or try to eliminate the currently adopted standards, although it wants them harmonised with the EU's policies.
This is why the CEN-CENELEC has established a joint technical committee on \ac{AI}\footnote{\url{https://www.cencenelec.eu/areas-of-work/cen-cenelec-topics/artificial-intelligence}} and defined a road map for \ac{AI} standardization \cite{CENroadMap} that includes the harmonization of IEC 61508 and other standards.
In fact, according to article 2(1)(c) of Regulation (EU) No 1025/2012, the CEN-CENELEC is the \ac{EU} authority for standards.
Nonetheless, we can see also that European countries start producing
guidelines and roadmaps \cite{germany20roadmap} on this subject.

So, given the very clear position of the proposed AI Act with respect to the possibility of using AI systems in particular critical infrastructures, we believe that the CEN-CENELEC, together with IEC will adapt IEC 61508, eventually opening to a safe use of AI systems also in critical infrastructures.
Importantly, the CEN-CENELEC \cite{CENroadMap} 
has already identified article 41 as a possible source of uncertainty in industry, given that it would explicitly cut out any non-harmonized IEC standard (i.e., IEC 61508).
In fact, article clearly 41.1 says that: \quotes{Where harmonised standards referred to in Article 40 do not exist or where the Commission considers that the relevant harmonised standards are insufficient or that there is a need to address specific safety or fundamental right concerns, the Commission may, by means of implementing acts, adopt common specifications in respect of the requirements set out in Chapter 2 of [Title III]. Those implementing acts shall be adopted in accordance with the examination procedure referred to in Article 74(2).}

Consequently, given all the aforementioned facts, we see an harmonization of IEC 61508 or its replacement by 2024, and this will open to at least one two  scenarios.
In the first scenario, we will have an opening to the use of AI systems in the context of critical infrastructures, whereas they can improve sustainability whilst guaranteeing safety.
While in a second scenario a very strict policy against AI systems in critical infrastructures will be maintained.

Again, as consequence of the analysis presented in this Section, we believe that this very first scenario is the most likely.
If that is correct, we envisage that a new stream of research on AI for critical infrastructures will be opening by the end of 2024, paving the way for AI systems to improve the sustainability of our society.
Nonetheless, it is important to stress that the use of AI does not come free of problems related to safety, fairness, transparency and sometimes even sustainability.
For this reason, in the following section we will discuss and classify existing AI techniques, to analyze their impact on sustainability and safety and to understand which AI-based solutions are likely to be allowed by a future harmonization of IEC 61508. 

\section{Classification and Discussion of the Impact of AI on Sustainability and Safety} \label{sec:classification}

As mentioned in Section \ref{sec:proposed_ai_act}, the techniques and approaches covered by the proposed AI Act include both symbolic (e.g., logic-based) and non-symbolic (e.g., neural networks, statistics) techniques.
Nonetheless, each different type of AI may have its own characteristics, impacting on safety differently from others.
Indeed, as suggested by Mohseni et al. in their taxonomy of machine learning safety \cite{mohseni2021SafetyML}, the decisions of state-of-the-art machine learning techniques can be, in some cases, completely unexplainable, non-transparent, biased and non-robust.
On the other hand, the automatic decisions of fully symbolic approaches can be explainable by design but not as good as those of a state-of-the-art neural network \cite{holzinger2018machine}.
Therefore, given such a trade-off between explainability and performance, being able to foresee and analyse the impact of AI on safety is not trivial, forcing us to analyse it differently for different types of applications.

This is why with the present paper we will study the impact of AI, on the safety and sustainability of critical infrastructures, by using as reference point the 4 \acl{SIL} defined by IEC 61508.
In fact, for each safety level we will show concrete examples of technological solutions based on AI that have the potential for significantly improving sustainability, analyzing what is the trade-off between sustainability and safety and how important that is.

\subsection{Examples of "Forbidden" AI-based Solutions that Could Improve Sustainability in Safety-Critical Systems}

Safety-critical subsystems of cyber-physical systems\footnote{A cyber-physical system comprises physical mechanisms that are monitored and/or controlled by Information and Communication Technologies.} compliant with IEC 61508 are required to have the properties listed in \Cref{whatIsIEC61508} and these normally do not include \ac{AI}.
Nonetheless, few examples can illustrates how impactful can be \ac{AI} on safety-critical systems, considering that in scientific and technical literature are available several studies that directly address the issue or propose promising approaches that well fit the kind of data relevant for safety-critical functions. 
In \Cref{sil_based_classification} we show these examples aligned to the \acl{SIL} of IEC 61508.

\begin{table}[!tb]
\centering
\caption{\textbf{AI on Safety-Critical Environments}: This table shows examples of possible applications of AI on some safety-critical contexts. For each context we identify its \acf{SIL} and possible tasks where AI can be deployed to improve sustainability.} \label{sil_based_classification}
\makebox[\linewidth]{
\begin{tabular}{|c|l|l|} 
\hline
\rowcolor[rgb]{0.937,0.937,0.937} \multicolumn{1}{|l|}{\textbf{SIL}} & \textbf{Context}            & \multicolumn{1}{c|}{\textbf{Use of AI}}                                                                                                                                           \\ 
\hline
4                                                                    & Nuclear power plant \cite{gomezFernandez20Nuclear}         & \begin{tabular}{@{\labelitemi\hspace{\dimexpr\labelsep+0.5\tabcolsep}}l@{}}Anomaly detection \cite{caliva18nuclearAnomaly,boza2019subtle}\\In-core full management \cite{nissan19nuclear}\end{tabular}                                                \\ 
\hline
3                                                                    & Railway, station management \cite{NLP16,IUC16} & \begin{tabular}{@{\labelitemi\hspace{\dimexpr\labelsep+0.5\tabcolsep}}l@{}}Turning on/off switch heaters \cite{chiaradonna2021RailwayEnergyConsumption}\\Fault detection of sensitive components \cite{VALU3S-21}\end{tabular}                           \\ 
\hline
2 \& 1                                                                 & Chemical industry \cite{Accenture14,WEF17}           & \begin{tabular}{@{\labelitemi\hspace{\dimexpr\labelsep+0.5\tabcolsep}}l@{}}Predicting chattering alarms \cite{tamascelli20chemicalPrediction}\\Plant health diagnosis \cite{hao20chemicals}\end{tabular}  \\
\hline
\end{tabular}
}
\end{table}

\acl{SIL} 4 systems compliant to IEC 61508 are quite rare.
Nevertheless, the nuclear power plants industry offers examples of such systems \cite{mahtinen10nuclear}.
Here, \ac{AI} is envisioned to have a great impact in the relatively close future,
in particular for safeguard and surveillance (filter and identify
signatures of nuclear materials), monitoring and diagnosis of severe
accidents or nuclear power plant transients \cite{nuclearInnovation19}.
All these actions are crucial to avoid environmental consequences
of accidents and lives lost \cite{nissan19nuclear,gomezFernandez20Nuclear}.

For \acl{SIL} 3 consider the railway industry, and in particular
those systems for which energy efficiency is crucial, 
with focus on the heating system for rail-road switches 
\cite{chiaradonna2021RailwayEnergyConsumption}. 
This is a critical subsystem, responsible for keeping the switches free from snow and ice, necessary to guarantee the correct operation of the switches and so the correct train routing (always turned on increases safety).  
Depending on the climate conditions of the place where the railway system operates, the energy consumed by this heating system can be very relevant, i.e. the heating always turned on implies greater ambient impact and cost. 
To provide concrete examples, in \cite{NLP16} it is reported that the cost for heating the $6800$ switches and crosses in Sweden can amount to $10-15$ Million Euros per year. 
In Germany, Deustche Bahn (DB) alone has $64000$ switches heated with electrical resistance and gas heaters, a combined power of 900 MW which consume up to 230 GWh/year \cite{IUC16}. 
\ac{AI} is expected to empower this subsystem with snow or ice 
prediction/detection and by making the turning on/off algorithm more responsive. 

Regarding \acl{SIL} 2 and 1, \cite{barone16SILoil} shown concrete examples
in the process industry, where the second tier standard IEC 61511 \cite{IEC61511}
apply, that can be generalized to the chemical industry.
The chemical industry is one of the most energy-intensive manufacturing industries and a major source of greenhouse gas emissions.
Besides that, chemical production often involves hazardous materials and high-pressure/high-temperature conditions, which may lead to fire, explosion, and other types of chemical accidents. Those chemical accidents could cause casualties, financial and social losses \cite{liao2022sustainability}.
According to a survey conducted by Accenture \cite{Accenture14}, most of the companies in the chemical and advanced
materials industry expect an industry-wide digitisation, and AI plays an essential role in enabling the digital revolution \cite{WEF17}.
In particular, fault detection and diagnosis is crucial to
both safety and sustainability.
As an example, consider fault detection for a Tennessee Eastman
process (chapter 8 of \cite{TennesseeEastmanProcess}) with few modes, where unit operations include
a reactor, a condenser, a recycle compressor, a vapour-liquid
separator and a stripper \cite{hao20chemicals}.  
Notice that the adoption of \ac{AI} is not `not recommended'
for \acl{SIL} 1 in IEC 61508.


Indeed, AI has the potential to cope with 
high dimensional data, being able to generalise, handling novel inputs and incomplete knowledge \cite{kurd07NNSafetyCritical}. 
These features are expected \cite{vinuesa2020role} to
greatly impact the way goals and targets in the 2030 Agenda for Sustainable Development are addressed.

Overall, we can say that \ac{AI} may be critical to anomaly detection, for taking timely countermeasures, being able to find patterns in data that do not conform to expected behaviour \cite{chandolaAnomaly09}.

\subsection{Discussion}
Even though several metrics for \ac{AI} performance and robustness appeared in literature and have been tested in several contexts \cite{wu2020testing}, only preliminary ones have been defined specifically to address safety or sustainability (e.g., \cite{bondavalliML19,bondavalliML21}), and are yet to be tested extensively before some \ac{AI} can become amenable for safety critical applications (where quantification has a central role).
Thus, it is expected that those \ac{AI} for which will be available reliable metrics will be the first to be employed in safety functions or safety critical systems.

It is desirable that interpretable or explainable-by-design \ac{AI} \cite{marcinkevics2020interpretability} 
are the first to be employed, in particular for handling tabular data \cite{rudin2019stop}.
This is indeed expected to cover, at least in part,
simplicity, understandability and observability-in-depth
(\Cref{safety_vs_AI}).

The heart of the problem is that \ac{AI} is difficult to be framed in safety standards because of the way it fails. 
Deterministic software fails systematically, whereas hardware fails randomly \cite{SafetyHandbook}.
Safety standards recommend to address hardware failures through statistical methods and mitigate/tolerate deterministic software failures employing qualitative techniques.
In some standards, statistical methods for quantifying software failures are allowed (e.g., suggestions are provided in Part 7 of IEC 61508 \cite{IEC61508}) in others (e.g., DO-178 C \cite{DO-178C}) are not recommended.
After about forty years of discussions, in industry and academia, no consensus has been reached, and strong opinions continue to emerge \cite{daniels22misuseStatistics}. 

Among those listed as \ac{AI} in Annex I of the proposed AI Act, some (e.g., statistical models or neural networks) are intrinsically non-deterministic \cite{johnson22AInondeterminism}, and then does not fit current safety standards framework.
Seen from a different perspective, though, this removes many of the assumptions that prevent the use of statistical methods,  opening up new ways to address \ac{AI} failures.
Indeed, a positive byproduct of the discussions on statistical methods for deterministic software is the huge body of knowledge that is available but not enough explored for addressing non-deterministic software.

\section{Conclusions} \label{sec:conclusion}

First of all, with this paper we performed an analysis of how the proposed AI Act might impact on the sustainability and safety of critical systems (e.g., power plants).
We did it by looking at the differences, incompatibilities and similarities of the AI Act with IEC 61508, one of the most important non-harmonised standards for safety-critical infrastructures.
Importantly, among the main differences, we show the incompatibility of IEC 61508 with the use of any AI in systems requiring a \acl{SIL} greater than 1, pointing to the disruptive effect that the proposed AI Act might have on that part of industry aligned with IEC 61508.
Then, we identified examples of AI-based solutions falling under the umbrella of IEC 61508 with a \acl{SIL} greater than 1 that might have a positive impact on sustainability in alignment with the current long-term goals of the EU and the proposed AI Act.


Eventually, we collected enough material to answer our initial research questions and foresee a future where critical infrastructures may harness the full potential of AI to improve both sustainability and safety in accordance with the following Sustainable Development Goals of the United Nations \cite{nations2016sustainable}: 
\begin{inparadesc}
    \item affordable and clean energy (goal 7),
    \item industry, innovation and infrastructure (goal 9),
    \item sustainable cities and communities (goal 11),
    \item responsible consumption and production (goal 12),
    \item climate action (13).
\end{inparadesc}
To be more precise, in accordance with the analysis we carried out in this paper, we believe that the AI Act will eventually soften the position of IEC 61508 with respect to AI, leading to a new generation of critical infrastructures harmonised with the European vision embodied by the proposed AI Act.
This would clearly open to new research and technological solutions on this topic by the end of 2024.
    

Overall, with this paper, our focus was exclusively on those safety-critical contexts where AI is expected to enhance economic/environmental aspects of sustainability but is not employed yet because considered not enough mature or potentially in conflict with safety or technical aspects of sustainability, as per IEC 61508.
Nonetheless, despite the promises made by state-of-the-art AI we can sceptically argue that using AI in safety-critical systems does definitely come with a risk.
This risk is posed by the fact that ceding control to machines might lead to new unregulated unethical and immoral behaviours as well as a dangerous lack of transparency and accountability.
Importantly, with respect to this specific issue, there are several flourishing discussions in literature and among policy makers, also taking into account that similar issues are addressed in other contexts as well  \cite{ghassemi2021MedicalExplainable}. This gives us hope that the technology of the future will be able to cope with such urgent problems to give us solutions based on AI capable of addressing the sustainability goals that have been set for the future.
For this reason, we argue that any forthcoming harmonised version of IEC 61508 is unlikely to completely close to application of AI in safety-critical systems with a \acl{SIL} greater than 1.
This is why we are all waiting for the CEN-CENELEC and its technical commission to give us a final answer to our research questions in the form of new harmonised standards.






\bibliographystyle{ACM-Reference-Format}
\bibliography{bib}


\begin{thebibliography}{52}


\ifx \showCODEN    \undefined \def \showCODEN     #1{\unskip}     \fi
\ifx \showDOI      \undefined \def \showDOI       #1{#1}\fi
\ifx \showISBNx    \undefined \def \showISBNx     #1{\unskip}     \fi
\ifx \showISBNxiii \undefined \def \showISBNxiii  #1{\unskip}     \fi
\ifx \showISSN     \undefined \def \showISSN      #1{\unskip}     \fi
\ifx \showLCCN     \undefined \def \showLCCN      #1{\unskip}     \fi
\ifx \shownote     \undefined \def \shownote      #1{#1}          \fi
\ifx \showarticletitle \undefined \def \showarticletitle #1{#1}   \fi
\ifx \showURL      \undefined \def \showURL       {\relax}        \fi
\providecommand\bibfield[2]{#2}
\providecommand\bibinfo[2]{#2}
\providecommand\natexlab[1]{#1}
\providecommand\showeprint[2][]{arXiv:#2}

\bibitem[\protect\citeauthoryear{??}{nat}{2016}]%
        {nations2016sustainable}
 \bibinfo{year}{2016}\natexlab{}.
\newblock \bibinfo{booktitle}{\emph{The Sustainable Development Goals Report}}.
\newblock \bibinfo{type}{{T}echnical {R}eport}. \bibinfo{institution}{United
  Nations}.
\newblock
\urldef\tempurl%
\url{https://unstats.un.org/sdgs/report/2016/The\%20Sustainable\%20Development\%20Goals\%20Report\%202016.pdf}
\showURL{%
\tempurl}


\bibitem[\protect\citeauthoryear{??}{CEN}{2020}]%
        {CENroadMap}
 \bibinfo{year}{2020}\natexlab{}.
\newblock \bibinfo{title}{Road Map on Artificial Intelligence (AI)}.
\newblock
\newblock
\urldef\tempurl%
\url{https://www.standict.eu/sites/default/files/2021-03/CEN-CLC_FGR_RoadMapAI.pdf}
\showURL{%
\tempurl}


\bibitem[\protect\citeauthoryear{167}{167}{1992}]%
        {DO-178C}
\bibfield{author}{\bibinfo{person}{RTCA (Firm).~SC 167}.}
  \bibinfo{year}{1992}\natexlab{}.
\newblock \bibinfo{booktitle}{\emph{Software considerations in airborne systems
  and equipment certification}}.
\newblock \bibinfo{publisher}{RTCA, Incorporated}.
\newblock


\bibitem[\protect\citeauthoryear{Accenture}{Accenture}{2016}]%
        {Accenture14}
\bibfield{author}{\bibinfo{person}{Accenture}.}
  \bibinfo{year}{2016}\natexlab{}.
\newblock \bibinfo{title}{Global Digital Chemistry - Survey quantitative
  findings}.
\newblock
\newblock
\urldef\tempurl%
\url{https://www2.deloitte.com/content/dam/Deloitte/de/Documents/consumer-industrial-products/Deloitte\%20Global\%20Digital\%20Chemistry\%20Survey2016Extract.pdf}
\showURL{%
\tempurl}


\bibitem[\protect\citeauthoryear{Avizienis, Laprie, Randell, and
  Landwehr}{Avizienis et~al\mbox{.}}{2004}]%
        {AvizienisTaxonomy}
\bibfield{author}{\bibinfo{person}{Algirdas Avizienis},
  \bibinfo{person}{Jean{-}Claude Laprie}, \bibinfo{person}{Brian Randell},
  {and} \bibinfo{person}{Carl~E. Landwehr}.} \bibinfo{year}{2004}\natexlab{}.
\newblock \showarticletitle{Basic Concepts and Taxonomy of Dependable and
  Secure Computing}.
\newblock \bibinfo{journal}{\emph{{IEEE} Trans. Dependable Secur. Comput.}}
  \bibinfo{volume}{1}, \bibinfo{number}{1} (\bibinfo{year}{2004}),
  \bibinfo{pages}{11--33}.
\newblock
\urldef\tempurl%
\url{https://doi.org/10.1109/TDSC.2004.2}
\showDOI{\tempurl}


\bibitem[\protect\citeauthoryear{Barbosa, Basagiannis, Giantamidis, Becker,
  Ferrari, Jahic, Kanak, Esnaola, Orani, Pereira, Pomante, Schlick, Smrcka,
  Yazici, Folkesson, and Sangchoolie}{Barbosa et~al\mbox{.}}{2020}]%
        {VALU3S-21}
\bibfield{author}{\bibinfo{person}{Raul Barbosa}, \bibinfo{person}{Stylianos
  Basagiannis}, \bibinfo{person}{Georgios Giantamidis}, \bibinfo{person}{H.
  Becker}, \bibinfo{person}{Enrico Ferrari}, \bibinfo{person}{J. Jahic},
  \bibinfo{person}{A. Kanak}, \bibinfo{person}{Mikel~Labayen Esnaola},
  \bibinfo{person}{Vanessa Orani}, \bibinfo{person}{David Pereira},
  \bibinfo{person}{Luigi Pomante}, \bibinfo{person}{Rupert Schlick},
  \bibinfo{person}{Ales Smrcka}, \bibinfo{person}{Ahmet Yazici},
  \bibinfo{person}{Peter Folkesson}, {and} \bibinfo{person}{Behrooz
  Sangchoolie}.} \bibinfo{year}{2020}\natexlab{}.
\newblock \showarticletitle{The {VALU3S} {ECSEL} Project: Verification and
  Validation of Automated Systems Safety and Security}. In
  \bibinfo{booktitle}{\emph{23rd Euromicro Conference on Digital System Design,
  {DSD} 2020, Kranj, Slovenia, August 26-28, 2020}}.
  \bibinfo{publisher}{{IEEE}}, \bibinfo{pages}{352--359}.
\newblock
\urldef\tempurl%
\url{https://doi.org/10.1109/DSD51259.2020.00064}
\showDOI{\tempurl}


\bibitem[\protect\citeauthoryear{Barone and Damiani}{Barone and
  Damiani}{2016}]%
        {barone16SILoil}
\bibfield{author}{\bibinfo{person}{D. Barone} {and} \bibinfo{person}{A.
  Damiani}.} \bibinfo{year}{2016}\natexlab{}.
\newblock \bibinfo{title}{Esperienza pratica nella applicazione delle analisi
  SIL (IEC 61508/61511) relative ai sistemi di sicurezza ad alta
  affidabilit{\`a}, per uno stabilimento a rischio di incidente rilevante}.
  (\bibinfo{year}{2016}).
\newblock
\urldef\tempurl%
\url{http://conference.ing.unipi.it/vgr2016/images/papers/133.pdf}
\showURL{%
\tempurl}
\newblock
\shownote{Valutazione e Gestione del Rischio negli Insediamenti Civili ed
  Industriali.}


\bibitem[\protect\citeauthoryear{Boza}{Boza}{2019}]%
        {boza2019subtle}
\bibfield{author}{\bibinfo{person}{Roger Boza}.}
  \bibinfo{year}{2019}\natexlab{}.
\newblock \bibinfo{booktitle}{\emph{Subtle Process Anomalies Detection using
  Machine Learning Methods}}.
\newblock \bibinfo{type}{{T}echnical {R}eport}. \bibinfo{institution}{U.S.
  Department of Energy, Office of Nuclear Energy}.
\newblock
\urldef\tempurl%
\url{https://lwrs.inl.gov/Advanced\%20IIC\%20System\%20Technologies/Subtle_Process-Anomalies_Detection_Using_Machine-Learning_Methods.pdf}
\showURL{%
\tempurl}


\bibitem[\protect\citeauthoryear{Caliv{\'{a}}, Ribeiro, Mylonakis,
  Demazi{\`{e}}re, Vinai, Leontidis, and Kollias}{Caliv{\'{a}}
  et~al\mbox{.}}{2018}]%
        {caliva18nuclearAnomaly}
\bibfield{author}{\bibinfo{person}{Francesco Caliv{\'{a}}},
  \bibinfo{person}{Fabio De~Sousa Ribeiro}, \bibinfo{person}{Antonios
  Mylonakis}, \bibinfo{person}{Christophe Demazi{\`{e}}re},
  \bibinfo{person}{Paolo Vinai}, \bibinfo{person}{Georgios Leontidis}, {and}
  \bibinfo{person}{Stefanos~D. Kollias}.} \bibinfo{year}{2018}\natexlab{}.
\newblock \showarticletitle{A Deep Learning Approach to Anomaly Detection in
  Nuclear Reactors}. In \bibinfo{booktitle}{\emph{2018 International Joint
  Conference on Neural Networks, {IJCNN} 2018, Rio de Janeiro, Brazil, July
  8-13, 2018}}. \bibinfo{publisher}{{IEEE}}, \bibinfo{pages}{1--8}.
\newblock
\urldef\tempurl%
\url{https://doi.org/10.1109/IJCNN.2018.8489130}
\showDOI{\tempurl}


\bibitem[\protect\citeauthoryear{Chandola, Banerjee, and Kumar}{Chandola
  et~al\mbox{.}}{2009}]%
        {chandolaAnomaly09}
\bibfield{author}{\bibinfo{person}{Varun Chandola}, \bibinfo{person}{Arindam
  Banerjee}, {and} \bibinfo{person}{Vipin Kumar}.}
  \bibinfo{year}{2009}\natexlab{}.
\newblock \showarticletitle{Anomaly Detection: A Survey}.
\newblock \bibinfo{journal}{\emph{ACM Comput. Surv.}} \bibinfo{volume}{41},
  \bibinfo{number}{3}, Article \bibinfo{articleno}{15} (\bibinfo{date}{jul}
  \bibinfo{year}{2009}), \bibinfo{numpages}{58}~pages.
\newblock
\showISSN{0360-0300}
\urldef\tempurl%
\url{https://doi.org/10.1145/1541880.1541882}
\showDOI{\tempurl}


\bibitem[\protect\citeauthoryear{Chiaradonna, Masetti, Giandomenico, Righetti,
  and Vallati}{Chiaradonna et~al\mbox{.}}{2021}]%
        {chiaradonna2021RailwayEnergyConsumption}
\bibfield{author}{\bibinfo{person}{Silvano Chiaradonna},
  \bibinfo{person}{Giulio Masetti}, \bibinfo{person}{Felicita~Di Giandomenico},
  \bibinfo{person}{Francesca Righetti}, {and} \bibinfo{person}{Carlo Vallati}.}
  \bibinfo{year}{2021}\natexlab{}.
\newblock \showarticletitle{Enhancing sustainability of the railway
  infrastructure: Trading energy saving and unavailability through efficient
  switch heating policies}.
\newblock \bibinfo{journal}{\emph{Sustain. Comput. Informatics Syst.}}
  \bibinfo{volume}{30} (\bibinfo{year}{2021}), \bibinfo{pages}{100519}.
\newblock
\urldef\tempurl%
\url{https://doi.org/10.1016/j.suscom.2021.100519}
\showDOI{\tempurl}


\bibitem[\protect\citeauthoryear{Commission}{Commission}{2021}]%
        {AIA}
\bibfield{author}{\bibinfo{person}{European Commission}.}
  \bibinfo{year}{2021}\natexlab{}.
\newblock \bibinfo{title}{Proposal for a Regulation of the European Parliament
  and of the Council laying down harmonised rules on Artificial Inteligence and
  amending certain union legislative acts}.
\newblock
\newblock
\urldef\tempurl%
\url{https://eur-lex.europa.eu/legal-content/EN/TXT/?uri=CELEX\%3A52021PC0206}
\showURL{%
\tempurl}


\bibitem[\protect\citeauthoryear{Commission, Centre, Tanarro~Colodron, Simola,
  Liessens, Joanny, Renda, Colle, Griveau, Vigier, Gerbelova, Vanleeuw, Cihlar,
  and Cambriani}{Commission et~al\mbox{.}}{2022}]%
        {nuclearInnovation19}
\bibfield{author}{\bibinfo{person}{European Commission},
  \bibinfo{person}{Joint~Research Centre}, \bibinfo{person}{J
  Tanarro~Colodron}, \bibinfo{person}{K Simola}, \bibinfo{person}{A Liessens},
  \bibinfo{person}{G Joanny}, \bibinfo{person}{G Renda}, \bibinfo{person}{J
  Colle}, \bibinfo{person}{J Griveau}, \bibinfo{person}{S Vigier},
  \bibinfo{person}{H Gerbelova}, \bibinfo{person}{D Vanleeuw},
  \bibinfo{person}{M Cihlar}, {and} \bibinfo{person}{A Cambriani}.}
  \bibinfo{year}{2022}\natexlab{}.
\newblock \bibinfo{booktitle}{\emph{Horizon scanning for nuclear safety and
  security yearly report 2021 : creating an anticipatory capacity within the
  JRC}}.
\newblock \bibinfo{publisher}{Publications Office of the European Union}.
\newblock
\urldef\tempurl%
\url{https://doi.org/doi/10.2760/645368}
\showDOI{\tempurl}


\bibitem[\protect\citeauthoryear{Daniels and Tudor}{Daniels and Tudor}{2022}]%
        {daniels22misuseStatistics}
\bibfield{author}{\bibinfo{person}{D. Daniels} {and} \bibinfo{person}{N.
  Tudor}.} \bibinfo{year}{2022}\natexlab{}.
\newblock \showarticletitle{Software Reliability and the Misuse of Statistics}.
\newblock \bibinfo{journal}{\emph{Safety-Critical Systems eJournal}}
  \bibinfo{volume}{1}, \bibinfo{number}{1} (\bibinfo{year}{2022}).
\newblock
\urldef\tempurl%
\url{https://scsc.uk/journal/index.php/scsj/article/view/8}
\showURL{%
\tempurl}


\bibitem[\protect\citeauthoryear{European Committee for Electrotechnical
  Standardization}{European Committee for Electrotechnical
  Standardization}{2020}]%
        {EN50128}
European Committee for Electrotechnical Standardization
  \bibinfo{year}{2020}\natexlab{}.
\newblock \bibinfo{booktitle}{\emph{Railway applications - Communication,
  signalling and processing systems - Software for railway control and
  protection systems}}.
\newblock European Committee for Electrotechnical Standardization.
\newblock


\bibitem[\protect\citeauthoryear{Evan L.~Russell}{Evan L.~Russell}{2000}]%
        {TennesseeEastmanProcess}
\bibfield{author}{\bibinfo{person}{Richard D.~Braatz Evan L.~Russell, Leo
  H.~Chiang}.} \bibinfo{year}{2000}\natexlab{}.
\newblock \bibinfo{booktitle}{\emph{Data-driven Methods for Fault Detection and
  Diagnosis in Chemical Processes}}.
\newblock
\urldef\tempurl%
\url{https://doi.org/10.1007/978-1-4471-0409-4}
\showDOI{\tempurl}


\bibitem[\protect\citeauthoryear{Forum}{Forum}{2017}]%
        {WEF17}
\bibfield{author}{\bibinfo{person}{World~Economic Forum}.}
  \bibinfo{year}{2017}\natexlab{}.
\newblock \bibinfo{title}{Digital transformation initiative chemistry and
  advanced materials industry}.
\newblock
\newblock
\urldef\tempurl%
\url{http://reports.weforum.org/digital-transformation/wp-content/blogs.dir/94/mp/files/pages/files/white-paper-dti-2017-chemistry.pdf}
\showURL{%
\tempurl}


\bibitem[\protect\citeauthoryear{Gharib and Bondavalli}{Gharib and
  Bondavalli}{2019}]%
        {bondavalliML19}
\bibfield{author}{\bibinfo{person}{Mohamad Gharib} {and}
  \bibinfo{person}{Andrea Bondavalli}.} \bibinfo{year}{2019}\natexlab{}.
\newblock \showarticletitle{On the Evaluation Measures for Machine Learning
  Algorithms for Safety-Critical Systems}. In \bibinfo{booktitle}{\emph{15th
  European Dependable Computing Conference, {EDCC} 2019, Naples, Italy,
  September 17-20, 2019}}. \bibinfo{publisher}{{IEEE}},
  \bibinfo{pages}{141--144}.
\newblock
\urldef\tempurl%
\url{https://doi.org/10.1109/EDCC.2019.00035}
\showDOI{\tempurl}


\bibitem[\protect\citeauthoryear{Gharib, Lollini, Botta, Amparore, Donatelli,
  and Bondavalli}{Gharib et~al\mbox{.}}{2018}]%
        {VetVforMLinISO26262}
\bibfield{author}{\bibinfo{person}{Mohamad Gharib}, \bibinfo{person}{Paolo
  Lollini}, \bibinfo{person}{Marco Botta}, \bibinfo{person}{Elvio~Gilberto
  Amparore}, \bibinfo{person}{Susanna Donatelli}, {and} \bibinfo{person}{Andrea
  Bondavalli}.} \bibinfo{year}{2018}\natexlab{}.
\newblock \showarticletitle{On the Safety of Automotive Systems Incorporating
  Machine Learning Based Components: {A} Position Paper}. In
  \bibinfo{booktitle}{\emph{48th Annual {IEEE/IFIP} International Conference on
  Dependable Systems and Networks Workshops, {DSN} Workshops 2018, Luxembourg,
  June 25-28, 2018}}. \bibinfo{publisher}{{IEEE} Computer Society},
  \bibinfo{pages}{271--274}.
\newblock
\urldef\tempurl%
\url{https://doi.org/10.1109/DSN-W.2018.00074}
\showDOI{\tempurl}


\bibitem[\protect\citeauthoryear{Gharib, Zoppi, and Bondavalli}{Gharib
  et~al\mbox{.}}{2021}]%
        {bondavalliML21}
\bibfield{author}{\bibinfo{person}{Mohamad Gharib}, \bibinfo{person}{Tommaso
  Zoppi}, {and} \bibinfo{person}{Andrea Bondavalli}.}
  \bibinfo{year}{2021}\natexlab{}.
\newblock \showarticletitle{Understanding the properness of incorporating
  machine learning algorithms in safety-critical systems}. In
  \bibinfo{booktitle}{\emph{{SAC} '21: The 36th {ACM/SIGAPP} Symposium on
  Applied Computing, Virtual Event, Republic of Korea, March 22-26, 2021}},
  \bibfield{editor}{\bibinfo{person}{Chih{-}Cheng Hung}, \bibinfo{person}{Jiman
  Hong}, \bibinfo{person}{Alessio Bechini}, {and} \bibinfo{person}{Eunjee
  Song}} (Eds.). \bibinfo{publisher}{{ACM}}, \bibinfo{pages}{232--234}.
\newblock
\urldef\tempurl%
\url{https://doi.org/10.1145/3412841.3442074}
\showDOI{\tempurl}


\bibitem[\protect\citeauthoryear{Ghassemi, Oakden-Rayner, and Beam}{Ghassemi
  et~al\mbox{.}}{2021}]%
        {ghassemi2021MedicalExplainable}
\bibfield{author}{\bibinfo{person}{M. Ghassemi}, \bibinfo{person}{L.
  Oakden-Rayner}, {and} \bibinfo{person}{A.~L. Beam}.}
  \bibinfo{year}{2021}\natexlab{}.
\newblock \showarticletitle{The false hope of current approaches to explainable
  artificial intelligence in health care}.
\newblock \bibinfo{journal}{\emph{The Lancet, digital health}}
  \bibinfo{volume}{3} (\bibinfo{year}{2021}), \bibinfo{pages}{E745--E750}.
\newblock
Issue 11.


\bibitem[\protect\citeauthoryear{Gibney et~al\mbox{.}}{Gibney
  et~al\mbox{.}}{2016}]%
        {gibney2016google}
\bibfield{author}{\bibinfo{person}{Elizabeth Gibney} {et~al\mbox{.}}}
  \bibinfo{year}{2016}\natexlab{}.
\newblock \showarticletitle{Google AI algorithm masters ancient game of Go}.
\newblock \bibinfo{journal}{\emph{Nature}} \bibinfo{volume}{529},
  \bibinfo{number}{7587} (\bibinfo{year}{2016}), \bibinfo{pages}{445--446}.
\newblock


\bibitem[\protect\citeauthoryear{Gomez-Fernandez, Higley, Tokuhiro, Welter,
  Wong, and Yang}{Gomez-Fernandez et~al\mbox{.}}{2020}]%
        {gomezFernandez20Nuclear}
\bibfield{author}{\bibinfo{person}{Mario Gomez-Fernandez},
  \bibinfo{person}{Kathryn Higley}, \bibinfo{person}{Akira Tokuhiro},
  \bibinfo{person}{Kent Welter}, \bibinfo{person}{Weng-Keen Wong}, {and}
  \bibinfo{person}{Haori Yang}.} \bibinfo{year}{2020}\natexlab{}.
\newblock \showarticletitle{Status of research and development of
  learning-based approaches in nuclear science and engineering: A review}.
\newblock \bibinfo{journal}{\emph{Nuclear Engineering and Design}}
  \bibinfo{volume}{359} (\bibinfo{year}{2020}), \bibinfo{pages}{110479}.
\newblock
\showISSN{0029-5493}
\urldef\tempurl%
\url{https://doi.org/10.1016/j.nucengdes.2019.110479}
\showDOI{\tempurl}


\bibitem[\protect\citeauthoryear{He, Zhang, Ren, and Sun}{He
  et~al\mbox{.}}{2015}]%
        {he2015delving}
\bibfield{author}{\bibinfo{person}{Kaiming He}, \bibinfo{person}{Xiangyu
  Zhang}, \bibinfo{person}{Shaoqing Ren}, {and} \bibinfo{person}{Jian Sun}.}
  \bibinfo{year}{2015}\natexlab{}.
\newblock \showarticletitle{Delving deep into rectifiers: Surpassing
  human-level performance on imagenet classification}. In
  \bibinfo{booktitle}{\emph{Proceedings of the IEEE international conference on
  computer vision}}. \bibinfo{pages}{1026--1034}.
\newblock


\bibitem[\protect\citeauthoryear{Henriksson, Borg, and Englund}{Henriksson
  et~al\mbox{.}}{2018}]%
        {GapISO26262andML}
\bibfield{author}{\bibinfo{person}{Jens Henriksson}, \bibinfo{person}{Markus
  Borg}, {and} \bibinfo{person}{Cristofer Englund}.}
  \bibinfo{year}{2018}\natexlab{}.
\newblock \showarticletitle{Automotive Safety and Machine Learning: Initial
  Results from a Study on How to Adapt the {ISO} 26262 Safety Standard}. In
  \bibinfo{booktitle}{\emph{1st {IEEE/ACM} International Workshop on Software
  Engineering for {AI} in Autonomous Systems, SEFAIAS@ICSE 2018, Gothenburg,
  Sweden, May 28, 2018}}, \bibfield{editor}{\bibinfo{person}{Reinhard Stolle},
  \bibinfo{person}{Stephan Scholz}, {and} \bibinfo{person}{Manfred Broy}}
  (Eds.). \bibinfo{publisher}{{ACM}}, \bibinfo{pages}{47--49}.
\newblock
\urldef\tempurl%
\url{https://doi.org/10.1145/3194085.3194090}
\showDOI{\tempurl}


\bibitem[\protect\citeauthoryear{Holzinger}{Holzinger}{2018}]%
        {holzinger2018machine}
\bibfield{author}{\bibinfo{person}{Andreas Holzinger}.}
  \bibinfo{year}{2018}\natexlab{}.
\newblock \showarticletitle{From Machine Learning to Explainable AI}. In
  \bibinfo{booktitle}{\emph{2018 World Symposium on Digital Intelligence for
  Systems and Machines (DISA)}}. \bibinfo{pages}{55--66}.
\newblock
\urldef\tempurl%
\url{https://doi.org/10.1109/DISA.2018.8490530}
\showDOI{\tempurl}


\bibitem[\protect\citeauthoryear{International Electrotechnical
  Commission}{International Electrotechnical Commission}{2006}]%
        {IEC60880}
International Electrotechnical Commission \bibinfo{year}{2006}\natexlab{}.
\newblock \bibinfo{booktitle}{\emph{Nuclear power plants – Instrumentation
  and control systems important to safety – Software aspects for
  computer-based systems performing category A functions}}.
\newblock International Electrotechnical Commission.
\newblock


\bibitem[\protect\citeauthoryear{International Electrotechnical
  Commission}{International Electrotechnical Commission}{2010}]%
        {IEC61508}
International Electrotechnical Commission \bibinfo{year}{2010}\natexlab{}.
\newblock \bibinfo{booktitle}{\emph{Functional safety of
  electrical/electronic/programmable electronic safety-related systems}}.
\newblock International Electrotechnical Commission.
\newblock


\bibitem[\protect\citeauthoryear{International Electrotechnical
  Commission}{International Electrotechnical Commission}{2016}]%
        {IEC61511}
International Electrotechnical Commission \bibinfo{year}{2016}\natexlab{}.
\newblock \bibinfo{booktitle}{\emph{Functional safety - Safety instrumented
  systems for the process industry sector}}.
\newblock International Electrotechnical Commission.
\newblock


\bibitem[\protect\citeauthoryear{International Electrotechnical
  Commission}{International Electrotechnical Commission}{2021}]%
        {IEC62061}
International Electrotechnical Commission \bibinfo{year}{2021}\natexlab{}.
\newblock \bibinfo{booktitle}{\emph{Safety of machinery - Functional safety of
  safety-related control systems}}.
\newblock International Electrotechnical Commission.
\newblock


\bibitem[\protect\citeauthoryear{Johnson}{Johnson}{2022}]%
        {johnson22AInondeterminism}
\bibfield{author}{\bibinfo{person}{Bonnie Johnson}.}
  \bibinfo{year}{2022}\natexlab{}.
\newblock \showarticletitle{Metacognition for artificial intelligence system
  safety – An approach to safe and desired behavior}.
\newblock \bibinfo{journal}{\emph{Safety Science}}  \bibinfo{volume}{151}
  (\bibinfo{year}{2022}), \bibinfo{pages}{105743}.
\newblock
\showISSN{0925-7535}
\urldef\tempurl%
\url{https://doi.org/10.1016/j.ssci.2022.105743}
\showDOI{\tempurl}


\bibitem[\protect\citeauthoryear{Kurd, Kelly, and Austin}{Kurd
  et~al\mbox{.}}{2007}]%
        {kurd07NNSafetyCritical}
\bibfield{author}{\bibinfo{person}{Zeshan Kurd}, \bibinfo{person}{Tim Kelly},
  {and} \bibinfo{person}{Jim Austin}.} \bibinfo{year}{2007}\natexlab{}.
\newblock \showarticletitle{Developing artificial neural networks for safety
  critical systems}.
\newblock \bibinfo{journal}{\emph{Neural Comput. Appl.}} \bibinfo{volume}{16},
  \bibinfo{number}{1} (\bibinfo{year}{2007}), \bibinfo{pages}{11--19}.
\newblock
\urldef\tempurl%
\url{https://doi.org/10.1007/s00521-006-0039-9}
\showDOI{\tempurl}


\bibitem[\protect\citeauthoryear{Lahtinen, Johansson, Ranta, Harju, and
  Nevalainen}{Lahtinen et~al\mbox{.}}{2010}]%
        {mahtinen10nuclear}
\bibfield{author}{\bibinfo{person}{Jussi Lahtinen}, \bibinfo{person}{Mika
  Johansson}, \bibinfo{person}{Jukka Ranta}, \bibinfo{person}{Hannu Harju},
  {and} \bibinfo{person}{Risto Nevalainen}.} \bibinfo{year}{2010}\natexlab{}.
\newblock \showarticletitle{Comparison between {IEC} 60880 and {IEC} 61508 for
  Certification Purposes in the Nuclear Domain}. In
  \bibinfo{booktitle}{\emph{Computer Safety, Reliability, and Security, 29th
  International Conference, {SAFECOMP} 2010, Vienna, Austria, September 14-17,
  2010. Proceedings}} \emph{(\bibinfo{series}{Lecture Notes in Computer
  Science})}, \bibfield{editor}{\bibinfo{person}{Erwin Schoitsch}} (Ed.),
  Vol.~\bibinfo{volume}{6351}. \bibinfo{publisher}{Springer},
  \bibinfo{pages}{55--67}.
\newblock
\urldef\tempurl%
\url{https://doi.org/10.1007/978-3-642-15651-9\_5}
\showDOI{\tempurl}


\bibitem[\protect\citeauthoryear{Liao, Lan, and Yao}{Liao
  et~al\mbox{.}}{2022}]%
        {liao2022sustainability}
\bibfield{author}{\bibinfo{person}{M. Liao}, \bibinfo{person}{K. Lan}, {and}
  \bibinfo{person}{Y. Yao}.} \bibinfo{year}{2022}\natexlab{}.
\newblock \showarticletitle{Sustainability implications of artificial
  intelligence in the chemical industry: A conceptual framework}.
\newblock \bibinfo{journal}{\emph{Journal of industrial ecology}}
  \bibinfo{volume}{26}, \bibinfo{number}{1} (\bibinfo{year}{2022}),
  \bibinfo{pages}{164--182}.
\newblock


\bibitem[\protect\citeauthoryear{Marcinkevics and Vogt}{Marcinkevics and
  Vogt}{2020}]%
        {marcinkevics2020interpretability}
\bibfield{author}{\bibinfo{person}{Ricards Marcinkevics} {and}
  \bibinfo{person}{Julia~E. Vogt}.} \bibinfo{year}{2020}\natexlab{}.
\newblock \showarticletitle{Interpretability and Explainability: {A} Machine
  Learning Zoo Mini-tour}.
\newblock \bibinfo{journal}{\emph{CoRR}}  \bibinfo{volume}{abs/2012.01805}
  (\bibinfo{year}{2020}).
\newblock
\showeprint[arXiv]{2012.01805}
\urldef\tempurl%
\url{https://arxiv.org/abs/2012.01805}
\showURL{%
\tempurl}


\bibitem[\protect\citeauthoryear{Mohseni, Wang, Yu, Xiao, Wang, and
  Yadawa}{Mohseni et~al\mbox{.}}{2021}]%
        {mohseni2021SafetyML}
\bibfield{author}{\bibinfo{person}{S. Mohseni}, \bibinfo{person}{H. Wang},
  \bibinfo{person}{Z. Yu}, \bibinfo{person}{C. Xiao}, \bibinfo{person}{Z.
  Wang}, {and} \bibinfo{person}{J. Yadawa}.} \bibinfo{year}{2021}\natexlab{}.
\newblock \bibinfo{title}{Taxonomy of Machine Learning Safety: A Survey and
  Primer}.
\newblock
\newblock
\urldef\tempurl%
\url{https://arxiv.org/abs/2106.04823}
\showURL{%
\tempurl}


\bibitem[\protect\citeauthoryear{Munro, Whiton, and Maxim}{Munro
  et~al\mbox{.}}{2019}]%
        {munro2019jobs}
\bibfield{author}{\bibinfo{person}{Mark Munro}, \bibinfo{person}{Jacob Whiton},
  {and} \bibinfo{person}{Robert Maxim}.} \bibinfo{year}{2019}\natexlab{}.
\newblock \showarticletitle{What jobs are affected by AI?}
\newblock  (\bibinfo{year}{2019}).
\newblock


\bibitem[\protect\citeauthoryear{Nissan}{Nissan}{2019}]%
        {nissan19nuclear}
\bibfield{author}{\bibinfo{person}{Ephraim Nissan}.}
  \bibinfo{year}{2019}\natexlab{}.
\newblock \showarticletitle{An Overview of AI Methods for in-Core Fuel
  Management: Tools for the Automatic Design of Nuclear Reactor Core
  Configurations for Fuel Reload, (Re)arranging New and Partly Spent Fuel}.
\newblock \bibinfo{journal}{\emph{Designs}} \bibinfo{volume}{3},
  \bibinfo{number}{3} (\bibinfo{year}{2019}).
\newblock
\showISSN{2411-9660}
\urldef\tempurl%
\url{https://doi.org/10.3390/designs3030037}
\showDOI{\tempurl}


\bibitem[\protect\citeauthoryear{of~Railways}{of~Railways}{[n. d.]}]%
        {IUC16}
\bibfield{author}{\bibinfo{person}{International~Union of Railways}.}
  \bibinfo{year}{[n. d.]}\natexlab{}.
\newblock \bibinfo{title}{Technologies and Potential Developments for Energy
  Efficiency and CO2 Reduction in Rail Systems}.
\newblock
\newblock
\urldef\tempurl%
\url{https://uic.org/IMG/pdf/_27_technologies_and_potential_developments_for_energy_efficiency_and_co2_reductions_in_rail_systems._uic_in_colaboration.pdf}
\showURL{%
\tempurl}
\newblock
\shownote{Online; accessed 15 January 2019.}


\bibitem[\protect\citeauthoryear{P.~Norbbin}{P.~Norbbin}{2016}]%
        {NLP16}
\bibfield{author}{\bibinfo{person}{A.~Parida P.~Norbbin, J.~Lin}.}
  \bibinfo{year}{2016}\natexlab{}.
\newblock \showarticletitle{Energy efficiency optimization for railway switches
  \& crossings: a case study in Sweden}. In \bibinfo{booktitle}{\emph{WCRR
  2016, 11th World Congress on Railway Research}}. \bibinfo{publisher}{SPARK
  knowledge sharing portal}.
\newblock
\urldef\tempurl%
\url{https://www.diva-portal.org/smash/get/diva2:1010747/FULLTEXT01.pdf}
\showURL{%
\tempurl}


\bibitem[\protect\citeauthoryear{Rajpurkar, Irvin, Zhu, Yang, Mehta, Duan,
  Ding, Bagul, Langlotz, Shpanskaya, et~al\mbox{.}}{Rajpurkar
  et~al\mbox{.}}{2017}]%
        {rajpurkar2017chexnet}
\bibfield{author}{\bibinfo{person}{Pranav Rajpurkar}, \bibinfo{person}{Jeremy
  Irvin}, \bibinfo{person}{Kaylie Zhu}, \bibinfo{person}{Brandon Yang},
  \bibinfo{person}{Hershel Mehta}, \bibinfo{person}{Tony Duan},
  \bibinfo{person}{Daisy Ding}, \bibinfo{person}{Aarti Bagul},
  \bibinfo{person}{Curtis Langlotz}, \bibinfo{person}{Katie Shpanskaya},
  {et~al\mbox{.}}} \bibinfo{year}{2017}\natexlab{}.
\newblock \showarticletitle{Chexnet: Radiologist-level pneumonia detection on
  chest x-rays with deep learning}.
\newblock \bibinfo{journal}{\emph{arXiv preprint arXiv:1711.05225}}
  (\bibinfo{year}{2017}).
\newblock


\bibitem[\protect\citeauthoryear{Rudin}{Rudin}{2019}]%
        {rudin2019stop}
\bibfield{author}{\bibinfo{person}{Cynthia Rudin}.}
  \bibinfo{year}{2019}\natexlab{}.
\newblock \showarticletitle{Stop explaining black box machine learning models
  for high stakes decisions and use interpretable models instead}.
\newblock \bibinfo{journal}{\emph{Nat. Mach. Intell.}} \bibinfo{volume}{1},
  \bibinfo{number}{5} (\bibinfo{year}{2019}), \bibinfo{pages}{206--215}.
\newblock
\urldef\tempurl%
\url{https://doi.org/10.1038/s42256-019-0048-x}
\showDOI{\tempurl}


\bibitem[\protect\citeauthoryear{Salay, Queiroz, and Czarnecki}{Salay
  et~al\mbox{.}}{2017}]%
        {salay2017MLinISO26262}
\bibfield{author}{\bibinfo{person}{Rick Salay}, \bibinfo{person}{Rodrigo
  Queiroz}, {and} \bibinfo{person}{Krzysztof Czarnecki}.}
  \bibinfo{year}{2017}\natexlab{}.
\newblock \showarticletitle{An Analysis of {ISO} 26262: Using Machine Learning
  Safely in Automotive Software}.
\newblock \bibinfo{journal}{\emph{CoRR}}  \bibinfo{volume}{abs/1709.02435}
  (\bibinfo{year}{2017}).
\newblock
\showeprint[arXiv]{1709.02435}
\urldef\tempurl%
\url{http://arxiv.org/abs/1709.02435}
\showURL{%
\tempurl}


\bibitem[\protect\citeauthoryear{Schrittwieser, Antonoglou, Hubert, Simonyan,
  Sifre, Schmitt, Guez, Lockhart, Hassabis, Graepel,
  et~al\mbox{.}}{Schrittwieser et~al\mbox{.}}{2020}]%
        {schrittwieser2020mastering}
\bibfield{author}{\bibinfo{person}{Julian Schrittwieser},
  \bibinfo{person}{Ioannis Antonoglou}, \bibinfo{person}{Thomas Hubert},
  \bibinfo{person}{Karen Simonyan}, \bibinfo{person}{Laurent Sifre},
  \bibinfo{person}{Simon Schmitt}, \bibinfo{person}{Arthur Guez},
  \bibinfo{person}{Edward Lockhart}, \bibinfo{person}{Demis Hassabis},
  \bibinfo{person}{Thore Graepel}, {et~al\mbox{.}}}
  \bibinfo{year}{2020}\natexlab{}.
\newblock \showarticletitle{Mastering atari, go, chess and shogi by planning
  with a learned model}.
\newblock \bibinfo{journal}{\emph{Nature}} \bibinfo{volume}{588},
  \bibinfo{number}{7839} (\bibinfo{year}{2020}), \bibinfo{pages}{604--609}.
\newblock


\bibitem[\protect\citeauthoryear{Sivageerthi, Sankaranarayanan, Ali, and
  Karuppiah}{Sivageerthi et~al\mbox{.}}{2022}]%
        {sivageerthi2022modelling}
\bibfield{author}{\bibinfo{person}{T Sivageerthi}, \bibinfo{person}{Bathrinath
  Sankaranarayanan}, \bibinfo{person}{Syed~Mithun Ali}, {and}
  \bibinfo{person}{Koppiahraj Karuppiah}.} \bibinfo{year}{2022}\natexlab{}.
\newblock \showarticletitle{Modelling the Relationships among the Key Factors
  Affecting the Performance of Coal-Fired Thermal Power Plants: Implications
  for Achieving Clean Energy}.
\newblock \bibinfo{journal}{\emph{Sustainability}} \bibinfo{volume}{14},
  \bibinfo{number}{6} (\bibinfo{year}{2022}), \bibinfo{pages}{3588}.
\newblock


\bibitem[\protect\citeauthoryear{Smith and Simpson}{Smith and Simpson}{2020a}]%
        {smith2020safety}
\bibfield{author}{\bibinfo{person}{David~J Smith} {and}
  \bibinfo{person}{Kenneth~GL Simpson}.} \bibinfo{year}{2020}\natexlab{a}.
\newblock \bibinfo{booktitle}{\emph{The Safety Critical Systems Handbook: A
  Straightforward Guide to Functional Safety: IEC 61508 (2010 Edition), IEC
  61511 (2015 Edition) and Related Guidance}}.
\newblock \bibinfo{publisher}{Butterworth-Heinemann}.
\newblock


\bibitem[\protect\citeauthoryear{Smith and Simpson}{Smith and Simpson}{2020b}]%
        {SafetyHandbook}
\bibfield{editor}{\bibinfo{person}{David~J. Smith} {and}
  \bibinfo{person}{Kenneth G.~L. Simpson}} (Eds.).
  \bibinfo{year}{2020}\natexlab{b}.
\newblock \bibinfo{booktitle}{\emph{The Safety Critical Systems Handbook}
  (\bibinfo{edition}{fifth edition} ed.)}.
\newblock


\bibitem[\protect\citeauthoryear{standards}{standards}{2020}]%
        {germany20roadmap}
\bibfield{author}{\bibinfo{person}{DKE standards}.}
  \bibinfo{year}{2020}\natexlab{}.
\newblock \bibinfo{title}{German standardization roadmap on artificial
  intelligence}.
\newblock
\newblock
\urldef\tempurl%
\url{https://www.din.de/resource/blob/772610/e96c34dd6b12900ea75b460538805349/normungsroadmap-en-data.pdf}
\showURL{%
\tempurl}


\bibitem[\protect\citeauthoryear{Tamascelli, Paltrinieri, and
  Cozzani}{Tamascelli et~al\mbox{.}}{2020}]%
        {tamascelli20chemicalPrediction}
\bibfield{author}{\bibinfo{person}{Nicola Tamascelli}, \bibinfo{person}{Nicola
  Paltrinieri}, {and} \bibinfo{person}{Valerio Cozzani}.}
  \bibinfo{year}{2020}\natexlab{}.
\newblock \showarticletitle{Predicting chattering alarms: {A} machine Learning
  approach}.
\newblock \bibinfo{journal}{\emph{Comput. Chem. Eng.}}  \bibinfo{volume}{143}
  (\bibinfo{year}{2020}), \bibinfo{pages}{107122}.
\newblock
\urldef\tempurl%
\url{https://doi.org/10.1016/j.compchemeng.2020.107122}
\showDOI{\tempurl}


\bibitem[\protect\citeauthoryear{Vinuesa, Azizpour, Leite, Balaam, Dignum,
  Domisch, Fell{\"{a}}nder, Langhans, Tegmark, and Nerini}{Vinuesa
  et~al\mbox{.}}{2019}]%
        {vinuesa2020role}
\bibfield{author}{\bibinfo{person}{Ricardo Vinuesa}, \bibinfo{person}{Hossein
  Azizpour}, \bibinfo{person}{Iolanda Leite}, \bibinfo{person}{Madeline
  Balaam}, \bibinfo{person}{Virginia Dignum}, \bibinfo{person}{Sami Domisch},
  \bibinfo{person}{Anna Fell{\"{a}}nder}, \bibinfo{person}{Simone Langhans},
  \bibinfo{person}{Max Tegmark}, {and} \bibinfo{person}{Francesco~Fuso
  Nerini}.} \bibinfo{year}{2019}\natexlab{}.
\newblock \showarticletitle{The role of artificial intelligence in achieving
  the Sustainable Development Goals}.
\newblock \bibinfo{journal}{\emph{CoRR}}  \bibinfo{volume}{abs/1905.00501}
  (\bibinfo{year}{2019}).
\newblock
\showeprint[arXiv]{1905.00501}
\urldef\tempurl%
\url{http://arxiv.org/abs/1905.00501}
\showURL{%
\tempurl}


\bibitem[\protect\citeauthoryear{W. and Z.}{W. and Z.}{2020}]%
        {hao20chemicals}
\bibfield{author}{\bibinfo{person}{Hao W.} {and} \bibinfo{person}{Jinsong Z.}}
  \bibinfo{year}{2020}\natexlab{}.
\newblock \showarticletitle{Fault detection and diagnosis based on transfer
  learning for multimode chemical processes}.
\newblock \bibinfo{journal}{\emph{Computers {\&} Chemical Engineering}}
  \bibinfo{volume}{135} (\bibinfo{year}{2020}), \bibinfo{pages}{106731}.
\newblock


\bibitem[\protect\citeauthoryear{Wu, Dong, Dong, Singa, Chen, and Zhang}{Wu
  et~al\mbox{.}}{2020}]%
        {wu2020testing}
\bibfield{author}{\bibinfo{person}{T. Wu}, \bibinfo{person}{Y. Dong},
  \bibinfo{person}{Z. Dong}, \bibinfo{person}{A. Singa}, \bibinfo{person}{X.
  Chen}, {and} \bibinfo{person}{Y. Zhang}.} \bibinfo{year}{2020}\natexlab{}.
\newblock \showarticletitle{Testing Artificial Intelligence System Towards
  Safety and Robustness: State of the Art}.
\newblock \bibinfo{journal}{\emph{International Journal of Computer Science}}
  \bibinfo{volume}{47}, \bibinfo{number}{3} (\bibinfo{year}{2020}).
\newblock


\end{thebibliography}

\end{document}